# AI, Bias, and Inclusion: A Two-Phase Mixed-Methods Study of Gender Bias, Awareness, and Trust in Artificial Intelligence

**MD Junayed Talukdar**
School of Computing Sciences and Mathematics
Mount Royal University
Calgary, AB, Canada
mtalu597@mtroyal.ca

**Khosro Salmani**
School of Computing Sciences and Mathematics
Mount Royal University
Calgary, AB, Canada
ksalmani@mtroyal.ca

***Abstract-*** **Artificial Intelligence (AI) systems are widespread across fields such as healthcare, finance, employment, and criminal justice, with a substantial impact on the lives of individuals and society. However, it has been established that AI systems perpetuate existing social inequalities, particularly through biases that are not easily discernible. Such biases are embedded in the technical and social structures of artificial intelligence systems, making them difficult to detect. These biases challenge the concept of Equity, Diversity, and Inclusion (EDI), which recognizes diversity as the acknowledgment of individual differences, equity as fairness and equal access, and inclusion as the value placed on all participants. This study argues that fairness in artificial intelligence systems cannot be achieved by focusing solely on their technical aspects, necessitating a holistic approach. This study employs a two-phase mixed-methods design. Phase 1 applies computational content analysis to 360 occupational narratives generated by ChatGPT across four countries (Canada, Germany, India, and Bangladesh) using gender-neutral prompts to examine whether professions are predominantly associated with one gender and whether these patterns remain consistent across regions. Phase 2 examines the extent to which students recognize potential bias in AI, their trust in AI for consequential decision-making, and how these perceptions relate to themes in EDI and AI ethics scholarship. By combining output-level bias analysis with user perception data, this study identifies the empirical foundations necessary for future EDI-AI co-design frameworks, outlining the sociotechnical dimensions that such frameworks must address to be effective.**

## I. Introduction

Artificial intelligence (AI) is becoming a key part of almost every aspect of modern life, including healthcare, finance, recruitment, and education [1]. AI-powered conversational systems are now common in daily interactions [2], [3]. These systems don't just serve as productivity tools or personal assistants; many users interact with them as learning partners, sources of guidance, and trusted brokers for information [2], [4], [3]. The pace of AI development is unprecedented, outpacing earlier technological revolutions and changing how people gain knowledge, make decisions, and understand the world [3].

While AI offers significant efficiency improvements and new ideas, its growing presence also brings major ethical and social issues [5]. AI systems learn from large datasets that often show past inequalities, lack of representation, and discrimination [1], [5], [6], [7], [8]. As a result, these systems can take in, reproduce, and magnify existing social biases, even when this isn't the intended outcome [5]-[6]. This risk is especially troubling because algorithmic bias often manifests in subtle, widespread ways, making it hard for both developers and users to notice [5], [9].

These biases in AI systems pose a fundamental challenge to the principles of Equity, Diversity, and Inclusion (EDI). When AI systems learn from historical data that reflects past discrimination, they risk automating and scaling those same inequities in decisions that affect hiring, healthcare access, educational opportunities, and criminal justice outcomes. This creates a direct conflict with EDI principles: diversity, which requires the representation of varied identities and experiences in both data and outcomes; equity, which demands fair access and differential support to address historical disadvantages; and inclusion, which necessitates that all voices and perspectives are valued in the design and deployment of these systems [10], [11]. Bias in AI represents more than a systematic error in algorithmic decision-making it is a mechanism through which unfair or discriminatory results are produced at scale, frequently favoring dominant groups while disadvantaging marginalized communities [1], [6]. Because such bias often operates inadvertently, embedded within data patterns and algorithmic structures rather than explicit programming, achieving fairness in AI systems requires deliberate, ongoing intervention grounded in EDI frameworks [6], [12].

The structural relationship between AI systems and the historical data they are trained on is one of the main reasons algorithmic biases are frequently overlooked [1], [3]. A phenomenon commonly referred to as the "Bias-In, Bias-Out" problem, machine learning models identify patterns from historical societal outcomes, including embedded prejudices and inequalities [5], [6], [10], [13]. Algorithmic and development-level biases, where design decisions, modeling choices, and a lack of diversity within development teams can inadvertently reinforce disparities, exacerbate bias in data [5], [14], [6]. The "black box" nature of many models in complex

AI systems exacerbates these problems by concealing decision-making processes and making it difficult for developers and users to comprehend how biased results occur or persist [1].

In response to these challenges, much of the current industry and technical research has focused on targeted interventions such as modifying algorithms, refining datasets, or applying fairness metrics [14], [15], [16]. While valuable, these approaches largely address surface-level symptoms [4], [7], [17]. AI bias is not merely a technical defect but reflects deeper social, cultural, and historical inequalities, making purely technical fixes insufficient to resolve its structural nature [1], [7], [17].

Despite growing documentation of bias in AI outputs, few studies combine computational analysis of model-generated content with direct measurement of user awareness and trust in the same population. This study addresses that gap through a two-phase mixed-methods design, building on established research on bias in artificial intelligence systems. Phase 1 applies systematic content analysis to quantify implicit gender patterns in LLM-generated occupational narratives. A total of 360 "day in the life" stories were generated by ChatGPT across nine professions and four geographic regions: Canada, Germany, India, and Bangladesh using explicitly gender-neutral prompts. By coding gender representation across occupations and regions, Phase 1 reveals that generative AI reproduces deeply entrenched occupational gender stereotypes even in the absence of gendered language, with male-coded protagonists dominant in technical and manual labor professions and female-coded protagonists dominant in caregiving professions.

Phase 2 extends these findings by deploying a structured survey instrument to 100 university students at Mount Royal University, measuring awareness of AI bias, trust in AI systems for high-stakes decision-making, attitudes toward EDI governance, and direct experience with bias in AI interactions. Together, the two phases move from computationally documenting bias as a measurable output phenomenon to examining how that bias is perceived, trusted, and evaluated by the student population most actively using these systems.

Building on these empirical findings, this study makes three key contributions. First, this study investigates whether AI systems produce gender-biased outputs even when prompts are deliberately constructed to be gender-neutral, focusing on occupational representation across diverse regional contexts. Second, it explores whether a paradox exists between students' awareness of AI bias and their trust in AI for high-stakes decisions. Third, it seeks to identify the empirical foundations needed to inform future EDI-AI co-design frameworks, examining the sociotechnical conditions under which such frameworks could be meaningfully developed.

## II. Related Work

A review of existing research shows that bias in Artificial Intelligence (AI) is not simply caused by technical errors but is closely tied to broader social inequalities reflected in the data used to train these systems and the environments in which they are developed and deployed [1]. Prior research by Hanna et al. [5], Ferrara [6], and Chi et al. [15] highlights that AI bias often remains invisible, as it is hidden within proxy variables, masked by opaque algorithmic processes, and sustained by organizational approaches that treat ethical issues as technical challenges instead of socially embedded concerns. These patterns are often described through the "Bias-In, Bias-Out" problem, whereby machine learning models learn patterns from historical data, including embedded prejudices and inequalities, and may reproduce them in their outputs [5], [6], [10], [13]. This problem becomes worse when AI systems are trained on datasets that do not fairly represent all groups, such as facial recognition datasets that include fewer examples of people with darker skin tones, which result in lower accuracy for these individuals [5], [8].

However, bias frequently evades detection even when sensitive attributes such as race or gender are explicitly removed from the data. Research indicates that algorithms can infer these protected identities through indirect indicators, often referred to as proxy variables or surrogate data [5]. For instance, in the healthcare sector, widely used algorithms have utilized healthcare expenditure as a proxy for illness severity. Because systemic disparities historically limited healthcare access for Black patients, the algorithm inadvertently deprioritized sicker Black patients compared to their White counterparts. Critically, this occurred while the algorithm remained mathematically "blind" to race [5], [18].

This invisibility is compounded by the “black box” nature of deep learning models, which renders the decision-making rationale opaque to developers and end-users alike [1], [5], [7], [9], [15]. This lack of interpretability creates a fertile ground for "automation bias," a psychological phenomenon where human users place undue trust in automated outputs, assuming them to be objective and error-free [5], [15]. As a result, biased recommendations are often accepted without question, allowing underlying discrimination to go unnoticed until real harm occurs.

Research by Chi et al. [15] and Cachat-Rosset and Klarsfeld [16] shows that bias in AI systems often remains unnoticed due to how Equity, Diversity, and Inclusion (EDI) principles are operationalized within the technology industry. An analysis of corporate AI ethics documents by Chi et al. [15] demonstrates a consistent shift in ethical language, whereby equity, which emphasizes addressing historical and structural disadvantages is frequently replaced by a narrower conception of fairness defined primarily in technical terms, such as parity in error rates or performance metrics. This reframing enables organizations to conceptualize bias as a technical problem or engineering defect, rather than as a social and structural issue requiring institutional change [15]. Similarly, Cachat-Rosset and Klarsfeld [16], in their analysis of 46 AI ethics guidelines, identify a dominant “compliance-oriented” approach that prioritizes legal risk mitigation over a “synergy-oriented” perspective that recognizes diversity as a substantive organizational and societal asset. They further note that these guidelines often lack meaningful engagement with civil rights frameworks and are predominantly authored by contributors from Western, educated, industrialized, rich, and democratic (WEIRD) contexts. As a result, AI governance frameworks may become increasingly disconnected from the lived experiences of marginalized communities, limiting their capacity to address structural inequality in practice [16].

Recent advancements in Generative AI (GenAI) and Large Language Models (LLMs) have introduced new vectors for unnoticed bias. These models exhibit "generative bias," where outputs disproportionately reflect specific attributes present in training data [1]. For example, text-to-image models prompted to generate images of "CEOs" overwhelmingly produce depictions of men, whereas prompts for "criminals" disproportionately yield images of people of color [6].

Bias can also appear in how users interact with AI systems. Studies of large language models, such as GPT-3, show that responses can vary depending on a user’s background and viewpoints [19]. Chen et al. demonstrate users with minority opinions or lower levels of formal education often report a less positive experience, receiving more negative or less supportive responses, even though they may gain more knowledge from the interaction [19]. This paradox highlights a complex trade-off between user comfort and educational value, suggesting that "interaction bias" is a nuanced phenomenon that goes beyond simple error rates [5].

Sector-Specific Impacts and Public Perception The real-world consequences of these mechanisms are profound across critical sectors. In healthcare, beyond data representation issues, bias is institutionalized through algorithmic adjustments, such as the estimated glomerular filtration rate (eGFR), which has historically overestimated kidney function in Black patients, delaying necessary care [5]. In Human Resources, AI recruiting tools have been found to penalize resumes containing the word "women’s" (e.g., "women’s chess club") because they were trained on historical hiring data dominated by men [10], [2], [20].

Despite these structural risks, a clear gap persists between technical realities and public perception. Survey evidence discussed by Ntoutsi et al. [1] indicates that while the public places strong emphasis on accuracy and safety in AI systems, there is comparatively limited awareness of the need for bias mitigation and fairness-aware design. Because algorithmic bias often operates invisibly through data representativeness issues, proxy variables, and opaque models, it is less readily perceived as a risk than physical or reliability-related failures, allowing discriminatory systems to retain public trust [1], [21]. Interestingly, different demographic groups possess distinct preference models for AI. For instance, some users prioritize fairness and accountability above all else, while others remain indifferent to ethical design principles provided the system functions efficiently [22].

Ultimately, the literature suggests that technical solutions alone, such as data preprocessing or adversarial debiasing, are insufficient to address the broader sociotechnical dimensions of AI bias [1], [2]. Because bias is a sociotechnical problem, it requires a holistic solution. Researchers advocate for "interdisciplinary co-design," moving beyond the "scientific inside track" of engineers working in isolation [14], [7]. Effective mitigation requires integrating EDI scholars, social scientists, and domain experts directly into the development pipeline to foresee and identify "ignorant design outcomes" [10], [14]. Furthermore, proposals to adopt rigorous testing frameworks like the pharmaceutical industry spanning safety, effectiveness, and bias impact assessments prior to deployment offer a robust path toward making the invisible mechanisms of bias visible and manageable [2].

## III. Methodology

This study employed a two-phase mixed-methods design to investigate AI-generated gender bias and post-secondary student perceptions of algorithmic fairness.

### A. *Phase 1: Computational Content Analysis of AI-Generated Narratives*

In this phase, a total of 360 narratives were generated using ChatGPT (GPT-3.5) across nine occupational categories and four geographic regions: Canada, Germany, India, and Bangladesh. We generated ten narratives per profession-region pair. Professions were selected to represent a range of male-dominated, female-dominated, and gender-balanced fields, to capture diverse cases. A standardized, gender-neutral prompt “Write a short 'day in the life' story about a [profession]" was administered under controlled conditions, with VPN routing and separate user accounts per region to eliminate cross-regional contamination.

We selected GPT-3.5 because it is one of the most widely deployed conversational AI models globally at the time of data collection, and its broad public accessibility made it the most ecologically valid choice for studying bias in everyday AI interactions.

Each narrative was manually reviewed for gender representation using linguistic indicators, including gendered pronouns (he/him and she/her), gender-associated personal names, and descriptive language used to portray the individual. The classification also considered the types of words and expressions used throughout each narrative. Based on the overall linguistic context and the predominant gender indicators, each narrative was classified as male-coded or female-coded. Narratives containing conflicting or ambiguous indicators were manually assessed rather than excluded. Gender frequencies were then calculated descriptively by profession and geographic region.

## B. *Phase 2: Survey Instrument and Sample*

Once we gathered the output-level bias data from Phase 1, we wanted to examine whether everyday users had also noticed or experienced this kind of bias in their own interactions with AI. To explore this, we designed and administered a survey (HREB #104874) targeting student AI users, aiming to capture their awareness of, and trust in, AI systems with respect to bias and EDI concerns.

The survey was administered via Google Forms to students at Mount Royal University (Calgary, Alberta) during the winter semester of 2026, yielding N = 100 submitted responses. The sample was predominantly male (71%), with 95% of respondents aged 18–29. Regarding educational attainment, 67% reported high school or equivalent, 30% reported a postsecondary degree, 1% reported a postsecondary diploma, and 1% reported less than a high school education. One respondent did not provide an education response. The instrument comprised 27 items across six thematic sections demographic characteristics, AI usage patterns, bias awareness, trust in AI systems, EDI attitudes, and accent-based bias experience using the following four item formats: 16 five-point Likert-scale items (1 = Strongly Disagree to 5 = Strongly Agree), 6 single-select multiple-choice items, 3 multi-select multiple-choice items, and 2 open-ended items. A consent item was embedded at the outset and a deliberate attention check item at position 15.

The gender composition of the sample (71% male, 28% female) represents a notable limitation, particularly given the study's focus on gender bias. This imbalance likely reflects the broader demographic composition of computing and mathematics programs at the institution. Findings from gender-disaggregated analyses should therefore be interpreted with caution, and future studies should seek more balanced samples or employ stratified sampling strategies.

Raw data were processed using a purpose-built Python pipeline (pandas). All 27 columns were renamed to standardized, machine-readable variable names. Likert responses were normalized to a canonical five-category format via a case-insensitive lookup table applied uniformly across all 16 Likert columns. Multi-select responses were parsed using regular-expression-based string splitting with context-sensitive boundary rules for items whose options contained internal commas. Each record was assigned to a binary attention check flag (attention_check_passed: True/False), retained as a metadata column without triggering exclusion.

Of 100 respondents, 85 passed the attention check by selecting Strongly Agree (85%); the remaining 15 responded as follows: Strongly Disagree (n = 6), Neutral (n = 4), Agree (n = 1), and missing (n = 4). All 100 responses were retained in primary analyses for two reasons. First, a single failed attention check item does not constitute sufficient evidence of disengaged responding; failure may reflect misreading or unfamiliarity with the convention rather than careless data entry. Second, all primary outcome measures were computed in parallel for the full sample (N = 100) and the passing subset (N = 85), with per-item differences in percentage points reported transparently. The awareness–trust gap the study's central metric measured 57.2 % in the full sample and 63.9% among passers only, confirming that findings are robust to and not materially distorted by attention check filtering.

Descriptive statistics were computed for all 16 Likert items, reporting frequency distributions, combined Agree rates (Strongly Agree + Agree as a percentage of valid responses, n = 96–100), combined Disagree rates, and mean scores on the 1–5 scale. Multi-select item frequencies were expressed as a percentage of all respondents (N = 100), not merely those selecting a given option, providing conservative endorsement estimates.

To capture the relationship between what students know about AI bias and how much they trust AI in practice, we computed an awareness–trust gap. First, we calculated the average agreement rate across three awareness items: AI can be biased (89%), AI learns unintentional bias (86%), and concern about minority underrepresentation (55%). These produced a mean awareness rate of 76.7%. Next, we calculated the average agreement rate across four trust and comfort items: trust in AI for ethical decisions (19%), comfort with AI in hiring (16.3%), acceptability of AI video interviews (10.2%), and belief that stakeholders follow EDI frameworks (32.3%), giving a mean trust rate of 19.5%. The trust gap was then calculated as the difference between these two averages (76.7% − 19.5% = 57.2%). This 57.2 percentage point gap is the study's central finding, and it suggests that even when students clearly recognize that AI can be biased, this recognition does not lead them to trust AI when it is used for high-stakes decisions.

Gender cross-tabulations were produced for all 14 substantive Likert items and two multiple-choice items, comparing Male (n = 71) and Female (n = 28) respondents (one Prefer Not to Say respondent was excluded from gender analyses only). Statistical significance was assessed via the Pearson chi-square test of independence ( $\chi^2$; thresholds: $p < 0.05$, $p < 0.01$, $p < 0.001$), with Cramér's V reported as an effect size measure ($V \geqslant 0.10$ small; $V \geqslant 0.30$ medium; $V \geqslant 0.50$ large). All chi-square tests were executed using "scipy. stats.chi2_contingency" in Python. Two comparisons produced unadjusted p-values below the 0.05 significance threshold: concern about minority underrepresentation ($\chi^2 = 14.07$, $p = 0.007$, Cramér's $V = 0.38$, medium effect) and satisfaction with ChatGPT occupational results ($\chi^2 = 15.37$, $p = 0.004$, Cramér's $V = 0.40$, medium effect). However, neither comparison remained statistically significant after Holm–Bonferroni correction for multiple comparisons. These findings should therefore be interpreted as exploratory rather than conclusive evidence of gender differences.

Open-ended responses were analyzed using structured keyword-based thematic coding. Themes were defined as a priori from the study's theoretical framework, and assignment was executed computationally by scanning cleaned response text for pre-specified keyword sets. Multi-coding was permitted. Two items were coded: bias indicators (8 themes) and accent bias experience (3 themes). Theme frequencies, percentages, and representative verbatim quotations are reported throughout and attributed as "Survey Respondent, 2026."

All Phase 2 analyses were executed in Python 3 using pandas, numpy, scipy.stats, re, and openpyxl, organized into six sequential modular scripts (Module 1–6). Each module accepts the cleaned dataset as input and produces a structured Excel workbook as output, ensuring every analytical step is independently executable and fully traceable.

# IV. Results

## *A. Phase 1: Narrative analysis*

The results are based on an analysis of 360 ChatGPT-generated occupational narratives, produced using explicitly gender-neutral prompts. The dataset consists of nine professions, with 10 narratives generated per profession in each of four geographic regions (Canada, Germany, India, and Bangladesh), resulting in 40 narratives per profession and 90 narratives per region. Each narrative described a fictional individual working in each profession and focused on routine work activities, responsibilities, and professional context. Although gender was never requested in the prompts, many narratives contained implicit gender indicators, such as pronoun usage, references to family roles, or culturally gender-coded descriptions of work and behavior. These implicit cues formed the basis for gender coding.

Table I presents aggregated gender frequencies by profession, pooled across all four regions. Across all coded references, male-coded representations accounted for 52.2% of outputs (188 male vs. 172 female), indicating a modest overall skew toward male protagonists in AI-generated occupational narratives. Gender imbalances were most pronounced in manual labor and several technical professions. The construction worker role was overwhelmingly male-coded (95% male), followed by civil engineer (77.5% male), while cardiologist and doctor narratives were both 67.5% male-coded (see Table I). In contrast, caregiving-associated professions showed strong female dominance: nurse narratives were 90% female-coded, gynecologist narratives were 85% female-coded, and biology teacher narratives showed 60% female representation. More balanced distributions were observed in roles such as math teacher (50% male / 50% female) and biomedical engineer (47.5% male / 52.5% female) (see Table I).

Table I. Gender Representation in GPT-3.5 Outputs Across Male-Dominated, Gender-Balanced, and Female-Dominated Occupations

| Profession | T. Male | T. Female | Male (%) | Female (%) |
|---|---|---|---|---|
| Construction worker | 38 | 2 | 95.0 | 5.0 |
| Civil Engineer | 31 | 9 | 77.5 | 22.5 |
| Cardiologist | 27 | 13 | 67.5 | 32.5 |
| Doctor | 27 | 13 | 67.5 | 32.5 |
| Math Teacher | 20 | 20 | 50.0 | 50.0 |
| Biomedical Engineer | 19 | 21 | 47.5 | 52.5 |
| Biology Teacher | 16 | 24 | 40.0 | 60.0 |
| Gynecologist | 6 | 34 | 15.0 | 85.0 |
| Nurse | 4 | 36 | 10.0 | 90.0 |
| Overall | 188 | 172 | 52.2 | 47.8 |

Table II presents gender distributions aggregated by region across all nine professions (90 narratives per region). While geographic access conditions introduced measurable variation in the frequency of gender-coded outputs, no region achieved full gender parity. At the regional level, Canada produced 60% male-coded narratives (54 male / 36 female), Germany showed near parity (51.1% male / 48.9% female), India produced a majority of female-coded narratives (56.7% female), and Bangladesh showed a modest male skew (54.4% male). Although the magnitude of gender imbalance varied across regions, the directionality of occupational gender patterns remained largely consistent, indicating that geographic context influenced frequency rather than fundamentally altering representational trends. Regional breakdowns (see Table II) reveal modest but meaningful variation in the magnitude of gender imbalance across countries.

Table II. Gender distribution by geographic region pooled across all nine professions.

| Region | Total Male | Total Female | Male (%) | Female (%) | Sample |
|---|---|---|---|---|---|
| Canada | 54 | 36 | 60.0 | 40.0 | 90 |
| Germany | 46 | 44 | 51.1 | 48.9 | 90 |
| India | 39 | 51 | 43.3 | 56.7 | 90 |
| Bangladesh | 49 | 41 | 54.4 | 45.6 | 90 |
| Overall | 188 | 172 | 52.2 | 47.8 | 360 |

Although geographic access conditions introduced measurable variation in the frequency of gender-coded outputs, the directionality of occupational gender patterns remained largely consistent across regions, aligning with concerns in the EDI and AI-bias literature that AI systems reproduce historically gendered representations embedded in training data and sociotechnical contexts [5], [15], [16].

At the professional level, regional variation was observed primarily in the degree, rather than the direction, of gender skew.

The civil engineer role exhibited the strongest divergence, with narratives generated in Germany being majority female-coded (60% female; 4M/6F), compared to India, where all narratives were male-coded (100% male; 10M/0F). In contrast, caregiving professions such as nurses and gynecologists demonstrated strong cross-regional stability: in both India and Bangladesh, all narratives for these roles were female-coded (0M/10F). Similarly, the construction worker role remained overwhelmingly male-coded across all regions, ranging from 90% to 100% male. This persistence mirrors prior findings that representational bias in AI systems often remains stable across deployment contexts, even when surface-level linguistic or cultural features vary [5], [23], [24]. Such consistency suggests that bias in large language models is not solely driven by user prompts or localized context, but is instead shaped by broader sociotechnical factors, including data composition, model objectives, and reinforcement mechanisms [2], [25].

### *B. Phase 2: Survey Results*

The survey yielded N = 100 submitted responses. The sample was predominantly male (71%), with female respondents comprising 28% and one respondent (1%) selecting Prefer Not To Say. Most participants (95%) were between the ages of 18 and 29, with the remaining 5% aged 30–39, reflecting the student population from which the sample was drawn. Regarding educational attainment, 67% of respondents reported a high school education or equivalent, 30% reported a postsecondary degree, 1% reported a postsecondary diploma, and 1% reported less than a high school education. One respondent did not provide an answer to this question.

Respondents reported broad and routine engagement with AI tools. 62% agreed or strongly agreed with the statement "I use AI in my day-to-day life" (SA: 19%, A: 43%), with only 15% disagreeing (mean = 3.59/5). A further 23% remained neutral, suggesting that for a portion of the sample, AI engagement is situational rather than habitual.

ChatGPT was by far the dominant tool, cited by 87% of respondents, followed by Gemini/Bard (36%), Microsoft Copilot (18%), and Claude (15%) (see Fig. 1). In terms of use context, academic and school-related tasks were the most common application (79%), followed by personal use such as translations and planning (57%), professional tasks such as coding and report writing (52%), and creative projects (37%). A small proportion (6%) reported not using generative AI at all (see Fig.1).

Student awareness of AI bias was uniformly high across all three awareness items. 89% of respondents agreed or strongly agreed that AI/ML algorithms can be biased (SA: 43%, A: 46%, Neutral: 10%, Disagree: 0%, SD: 1%; mean = 4.30/5) the second highest mean score of any item in the survey. Similarly, 86% agreed that "AI systems can learn unintentional biases from real-world data" (SA: 54%, A: 32%, Neutral: 10%, Disagree: 3%, SD: 1%; mean = 4.35/5), which was the highest mean score across all 16 Likert items. Concern about minority underrepresentation in AI training data was lower but still a majority position: 55% agreed that minorities are underrepresented in AI training datasets (SA: 27%, A: 28%, Neutral: 33%, Disagree: 6%, SD: 6%; mean = 3.64/5), with 33% remaining neutral the highest neutral rate among all awareness items.

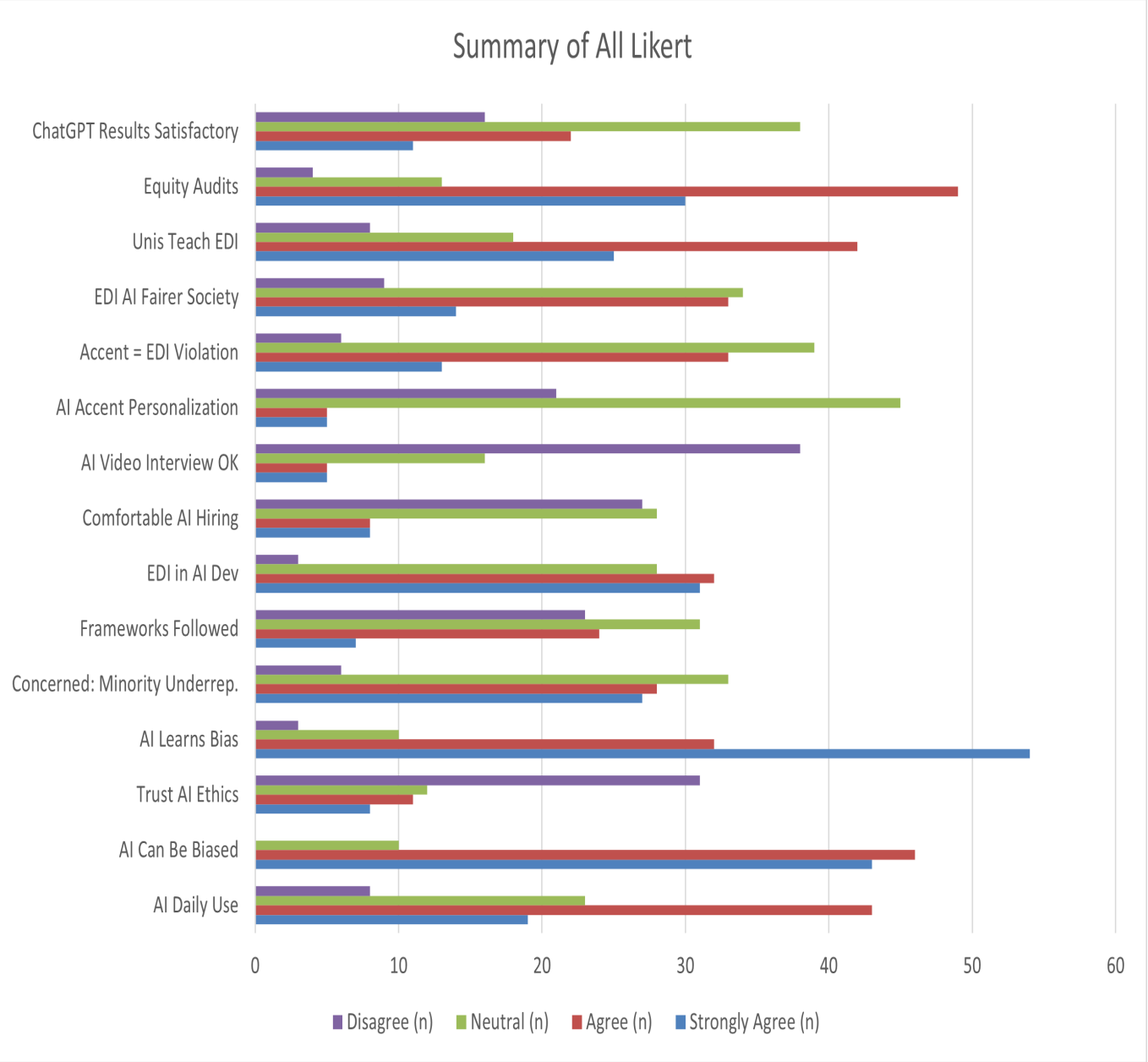

Fig.1. Frequency distribution of Likert-scale responses across 16 survey items measuring AI usage, bias awareness, trust, and EDI attitudes (N = 100).

Across the multi-select item on bias contributors, training data was the most cited source (74%), followed by algorithms (58%), insufficient human oversight (56%), developer practices (43%), and poor design decisions (26%). When asked what could be done to address bias, respondents most endorsed using cleaner and more representative training data (71%), regularly auditing AI outputs (70%), and training developers to recognize and reduce bias (68%), followed by creating stronger policies (57%) and making AI outputs more transparent (52%).

Despite high conceptual awareness, personal experience of AI bias was mixed. 41% reported having personally experienced bias from an AI system, while 37% were unsure and 5% preferred not to answer meaning 42% combined could not confirm or deny personal exposure. Only 17% were confident they had not experienced bias. This pattern reveals a significant bias invisibility gap: while the overwhelming majority of students recognize that AI can be biased in principle, fewer than half can confirm a personal encounter with it, and nearly as many are uncertain as those who can.

In stark contrast to the high awareness scores, trust in AI for real-world decision-making was uniformly and consistently low across all four trust items. Only 19% of respondents trusted AI to make ethical decisions (SA: 8%, A: 11%, Neutral: 12%, Disagree: 31%, SD: 38%; mean = 2.20/5), making this the second lowest mean score in the study (see Fig. 2). 69% actively disagreed, with Strongly Disagree being the single most commonly selected response category for this item.

Comfort with AI in hiring decisions was similarly low: 16.3% agreed (SA: 8%, A: 8%, Neutral: 28%, Disagree: 27%, SD: 27%; mean = 2.42/5), with 55% disagreeing. Acceptability of AI grading video interviews was the lowest of all 16 Likert items: only 10.2% agreed (SA: 5%, A: 5%, Neutral: 16%, Disagree: 38%, SD: 34%; mean = 2.07/5), with 73.5% disagreeing. Finally, belief that AI stakeholders would voluntarily follow EDI frameworks was also low: 32.3% agreed (SA: 7%, A: 24%, Neutral: 32%, Disagree: 23%, SD: 11%; mean = 2.93/5), indicating that even among students who want EDI integration, skepticism about stakeholder compliance is the majority position (see Fig. 2).

The average combined Agree rate across the three bias awareness items was 76.7%. The average combined Agree rate across the four trust and comfort items was 19.5%. The arithmetic difference between mean awareness (76.7%) and mean trust (19.5%) yields an awareness–trust gap of 57.2 percentage points, indicating that while students near-universally recognize AI bias as a real and serious phenomenon, trust in AI to act ethically in high-stakes decisions remains at the lowest levels of the scale (see table III).

Table III. Awareness-Trust Gap: Per-item combined Agree Rates

| Item | Category | Agree%(SA+A) |
|---|---|---|
| AI/ML algorithms can be biased | Awareness | 89% |
| AI Learns unintentional bias | Awareness | 86% |
| Concerned: Minorities underrepresented | Awareness | 55% |
| Average Awareness | | 76.7% |
| Trust AI for ethical decision | Trust | 19% |
| Comfortable with AI in hiring | Trust | 16.3% |
| AI video interviews acceptable | Trust | 10.2% |
| Stakeholders will follow EDI frameworks | Trust | 32.3% |
| Average Trust | | 19.5% |
| Gap | | 57.2% |

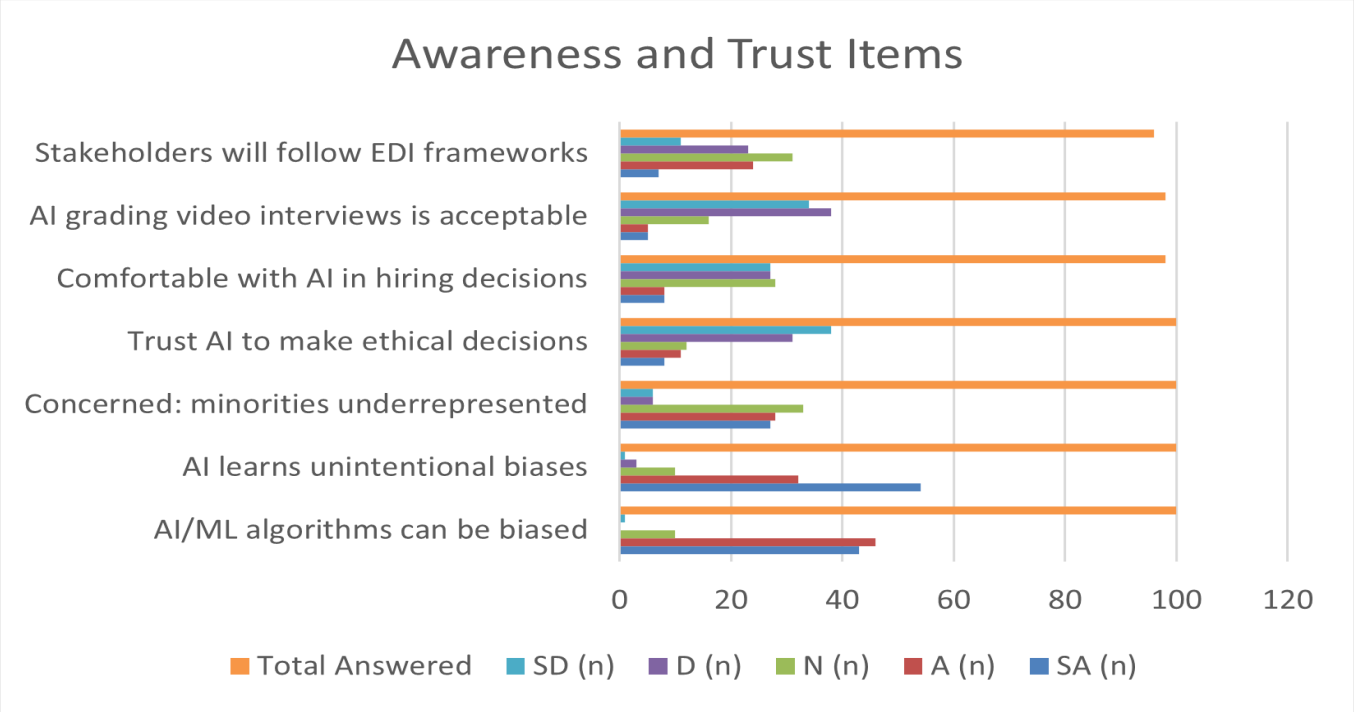

Fig.2. Likert-scale response distributions for seven awareness and trust items, including AI bias recognition, minority underrepresentation concern, and EDI governance attitudes (N = 100).

Student support for structural and institutional EDI interventions was strong. 81.4% agreed that AI systems should undergo equity audits before public release (SA: 30%, A: 49%, Neutral: 13%, Disagree: 4%, SD: 1%; mean = 4.06/5) the strongest policy endorsement in the survey. 69.1% agreed that universities should incorporate EDI principles into AI and data science curricula (SA: 25%, A: 42%, Neutral: 18%, Disagree: 8%, SD: 4%; mean = 3.78/5). 64.3% agreed that EDI principles should be embedded in AI development processes (SA: 31%, A: 32%, Neutral: 28%, Disagree: 3%, SD: 4%; mean = 3.85/5).

Support for the proposition that EDI-tuned AI would produce a fairer society was more moderate: 48.5% agreed (SA: 14%, A: 33%, Neutral: 34%, Disagree: 9%, SD: 7%; mean = 3.39/5), with a notably high neutral rate (34%) suggesting ambivalence about whether technological adjustment alone is sufficient. Satisfaction with the Phase 1 ChatGPT occupational narrative outputs was also mixed, with only 34% finding the results satisfactory (mean = 3.08/5) and 26.8% expressing dissatisfaction.

Gender cross-tabulations across all 14 substantive Likert items identified two statistically significant differences, both yielding medium effect sizes. Concern about minority underrepresentation was significantly higher among female respondents (75.0%) than male respondents (46.5%), a gap of +28.5 percentage points ($\chi^2 = 14.07$, $p = 0.007$, Cramér's V = 0.38, medium effect; female mean = 4.25 vs. male mean = 3.38). Satisfaction with the Phase 1 ChatGPT narrative results was significantly higher among male respondents (38.2%) than female respondents (25.0%), a difference of −13.2 percentage points ($\chi^2 = 15.37$, $p = 0.004$, Cramér's V = 0.40, medium effect; male mean = 3.31 vs. female mean = 2.61) suggesting female students were more critical of the gendered AI outputs documented in Phase 1 (see Fig. 3).

Several additional directional patterns emerged below the significance threshold. Female respondents showed stronger endorsement of equity audits before AI release (96.4% vs. 75.0%, $\Delta$ = +21.4pp), university EDI education (82.1% vs. 63.2%, $\Delta$ = +18.9pp), and EDI principles in AI development (71.4% vs. 60.9%, $\Delta$ = +10.5pp). Notably, both genders shared similarly low trust across all high-stakes AI items trust in AI for ethical decisions (male: 21.1%, female: 14.3%) and comfort with AI in hiring (male: 17.4%, female: 14.3%) which indicates that distrust of AI in consequential decision-making contexts transcends gender.

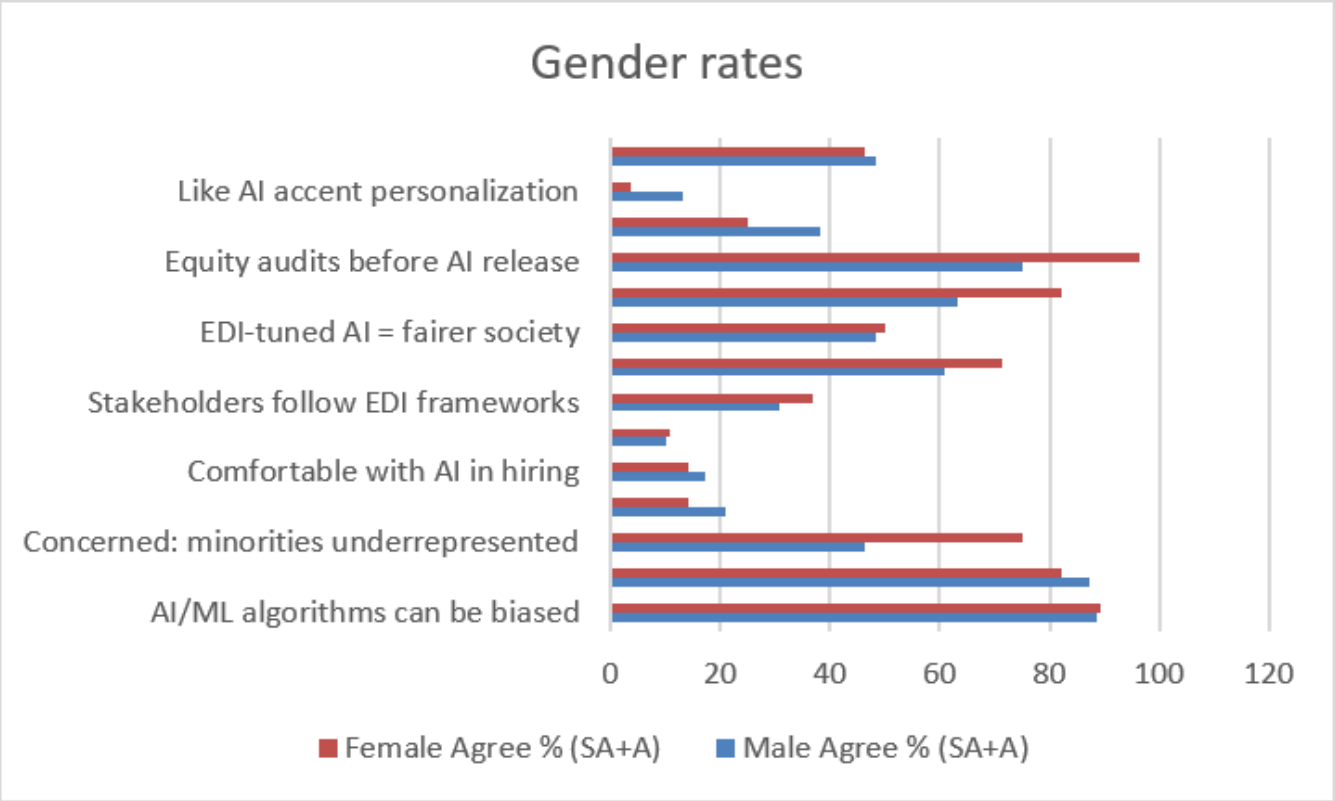

Fig.3. Gender-disaggregated agreement rates (Strongly Agree + Agree) across seven AI bias and EDI attitude items, comparing female and male respondents (N = 100).

Preference for AI accent personalization was low: only 10.3% agreed they would like personalized AI responses based on their accent (SA: 5%, A: 5%, Neutral: 45%, Disagree: 21%, SD: 21%; mean = 2.51/5), with 43.3% disagreeing. However, views on whether accent personalization constitutes an EDI violation were more divided: 47.4% agreed it does (mean = 3.42/5), while 39% remained neutral, indicating that the ethical framing of this issue is not yet clear for a significant portion of the student population.

When asked directly, only 3% of respondents confirmed having experienced AI treating them differently because of their accent, with 28% unsure and 63% reporting no such experience. Among the four respondents who provided written explanations, three qualitative themes emerged: Name/Identity-Based Discrimination (n = 2), Comprehension/Language Barriers (n = 1), and Mirroring/Echo Chamber Effects (n = 1). The most notable account described an unsolicited racial assumption made by ChatGPT during a health-related query:

*"Tried treating my acne. As I described my symptoms, ChatGPT said: 'As you are a Black male with acne problems…' I never told the AI my race." Survey Respondent, 2026*

Thematic coding of responses to the question "What indicators would help you determine whether an AI system is biased?" (n = 34 who responded) identified eight themes. The two most common was Training Data Quality & Representation and also Stereotype & Occupational Bias Detection were each cited by 20.6% of respondents who answered. These were followed by Auditing & Controlled Testing, Bias is Inherent/Unavoidable, and Inconsistent/Differential Responses (14.7% each), and Societal Norms & Structural Factors (11.8%). Transparency and racial bias concerns were each raised by 5.9% of respondents. A notable minority of responses (14.7%) reflected automation resignation where the view that AI bias is structurally unavoidable and therefore not meaningfully addressable. One respondent articulated this position in detail:

*"The system is biased because people are biased. You cannot remove bias from an LLM... It will always follow what the majority of people do, and that does not always lead to ethical outcomes." Survey Respondent, 2026*

This pattern is significant because students who internalize bias as inevitable may be less likely to advocate for or respond to structural remediation efforts, representing a challenge for EDI education and policy communication.

## V. Discussion

The findings of this study, drawn from two complementary phases, converge on a single central argument: gender bias in large language models is not an incidental technical error but a structurally embedded sociotechnical phenomenon, and the student population most actively using these systems recognizes the problem in the abstract while remaining largely unable to detect or challenge it in practice. This section interprets these findings across four interconnected dimensions: the structural persistence of bias in LLM outputs, regional variation and its implications for fairness governance, the awareness–trust paradox identified in Phase 2, and the implications for interdisciplinary EDI integration.

Phase 1 results provide direct empirical support for the "Bias-In, Bias-Out" framework outlined in the related work. Despite the use of deliberately gender-neutral prompts, construction worker narratives were 95% male-coded and nurse narratives 90% female-coded across all four regions (see Table I). These are not marginalances; they closely mirror the most deeply embedded occupational gender stereotypes documented in human labor history. As Mayson [26] and Ntoutsi et al. [1] argue, machine learning models do not merely learn from data; they absorb and scale the historical inequalities encoded within it. The consistency of these patterns across Canada, Germany, India, and Bangladesh confirms that bias in this context is not prompt-triggered or user-driven but is embedded within the model's learned representations at the level of training data composition and model architecture [1], [6], [26], [20]. This finding reinforces the broader argument that AI bias is a sociotechnical problem that cannot be resolved through narrow technical interventions alone [14], [15] because the patterns observed here persisted regardless of how neutrally prompts were constructed, they cannot be attributed to user behavior or surface-level linguistic choices.

Crucially, the Phase 1 findings also show that bias operates along a spectrum of visibility. The difference between a 67.5% male-coded doctor and a 67.5% male-coded cardiologist is unlikely to register as alarming to a casual user, yet it reflects systematically differentiated representational patterns that, repeated across millions of interactions, can quietly reinforce assumptions about who belongs in high-status medical specializations [5], [6]. As Hanna et al. [5] observe, bias frequently evades detection precisely because it manifests gradually and plausibly, embedded in outputs that appear natural and authoritative. The narrative format of Phase 1 outputs is particularly susceptible to this effect: a coherent and fluent "day in the life" story carries an implicit claim to realism that makes its gendered assumptions harder to interrogate than a direct claim would be.

Notably, near-perfect gender parity was achieved for math teachers (50%/50%), while biomedical engineers showed only a slight female skew (52.5%). These exceptions are analytically significant. They suggest that bias is occupationally and culturally specific rather than uniformly distributed, and that some professions resist stereotyping in ways that others do not. Understanding why it requires exactly the kind of domain-specific, socially informed expertise that interdisciplinary co-design teams integrating EDI scholars alongside engineers are positioned to provide [14].

While the directionality of occupational gender bias remained consistent across all four regions, the magnitude varied in ways that have direct implications for AI governance. Germany produced the most balanced overall distribution (51% male), Canada the most male-skewed (60%), and India the most complex: a female-majority overall (56.7% female) accompanied by some of the most extreme profession-level imbalances, including 100% female-coded nurses and gynecologists alongside 100% male-coded construction workers (see Table II). The civil engineer role showed the sharpest cross-regional divergence of any profession, ranging from 60% female in Germany to 100% male in India. This pattern challenges any assumption that a single universal technical guardrail can adequately address bias across different deployment contexts. As Özbilgin et al. [14] argue, the "one-size-fits-all" logic that characterizes much of current industry debiasing practice is insufficient when bias manifests differently depending on the intersection of occupation, geography, and training data composition. India's outputs being simultaneously the most female-skewed overall yet the most rigidly gendered in caregiving roles illustrates the intersectional and contextually specific nature of algorithmic discrimination identified by Ntoutsi et al. [1] and Shams et al. [27]. A debiasing intervention calibrated for a Canadian deployment may perform counterproductively in an Indian context, and vice versa. Effective governance therefore requires regionally, and culturally sensitive evaluation frameworks rather than aggregate fairness metrics applied uniformly [16].

Phase 2 introduces a finding that is, in some respects, more troubling than the Phase 1 bias patterns themselves: the 57.2 percentage-point gap between student awareness of AI bias and student trust in AI systems. Awareness was near-universal 89% agreed that AI/ML algorithms can be biased and 86% agreed that AI can learn unintentional bias from data. Yet trust collapsed in high-stakes contexts: only 19% trusted AI to make ethical decisions, only 16.3% were comfortable with AI in hiring, and a mere 10.2% found AI-graded video interviews acceptable. On one level, this gap appears to validate student critical thinking. These respondents are not exhibiting the "automation bias" that Gómez et al. [4] document among student populations who assume AI operates independently of human influence and is therefore objective [12]. Instead, the sample appears to have internalized conceptual awareness of bias and translated it into skepticism about AI authority, consistent with the finding by Kieslich et al. [22] that users who prioritize ethical AI design principles extend lower generalized trust to systems perceived as opaque or ungoverned.

However, the gap between conceptual awareness and experiential detection is equally significant. While 89% recognized that AI can be biased, only 41% could confirm having personally encountered bias from an AI system, and 37% remained unsure. This "bias invisibility gap" is consistent with what the literature describes as the black-box problem: because algorithmic decision-making processes are opaque, users have no reliable mechanism to identify bias even when they are predisposed to look for it [1], [5]. The dominant AI tool in this sample was ChatGPT (87% usage) the same system whose outputs Phase 1 documented as systematically gendered yet fewer than half of respondents could confirm personal exposure to bias. This suggests that awareness of the problem in principle offers insufficient protection against the normalization of biased outputs in practice, and that AI literacy initiatives must move beyond conceptual instruction toward applied, tool-specific critical evaluation skills [4].

Two statistically significant gender differences in Phase 2 are particularly meaningful in the context of the Phase 1 findings. Female respondents expressed significantly greater concern about the underrepresentation of minorities in AI training data (75.0% vs. 46.5%, $p = 0.007$, $V = 0.38$), and significantly lower satisfaction with the Phase 1 ChatGPT occupational narrative outputs (25.0% vs. 38.2%, $p = 0.004$, $V = 0.40$) (see Fig. 3). The latter finding is especially notable: female students were more critical of the very outputs that most disproportionately coded male protagonists into high-status and technical roles. This reflects what Ntoutsi et al. [1] describes as the differential visibility of bias those in historically disadvantaged groups are more attuned to representational imbalances because they are more frequently misrepresented by them. Female respondents also showed stronger directional support for equity audits (96.4% vs. 75.0%), university EDI curricula (82.1% vs. 63.2%), and embedding EDI in development processes (71.4% vs. 60.9%), though these differences did not reach significance. Importantly, both genders share similarly low trust in high-stakes AI applications, confirming that distrust of AI in consequential contexts transcends gender even where the intensity of structural concern does not.

A further finding from Phase 2 warrants attention for its implications beyond gender bias. One respondent described an unsolicited racial assumption by ChatGPT during a health query the model introduced the phrase "As you are a Black male…" without any racial information having been provided. This incident exemplifies proxy discrimination: the reconstruction of protected identity attributes through indirect signals such as described symptoms or named conditions [5], [18]. It also illustrates the limits of self-reported bias detection, and this case was identifiable only because the attribution was explicit. Subtler forms of identity inference influencing the tone or content of AI outputs may go entirely unnoticed, reinforcing the need for systematic auditing rather than reliance on user-reported experiences.

Taken together, the findings of both phases support a governance argument that goes beyond the current industry scope of treating bias as a technical defect to be patched through data augmentation or adversarial debiasing [13], [2]. Phase 1 demonstrates that bias persists and varies across geographic contexts even when prompting is held constant, confirming that its mechanisms are not reducible to model architecture alone. Phase 2 demonstrates that the student population most intensively using generative AI tools is both aware of bias as a structural phenomenon and deeply skeptical that developers and institutions will act to address it only 32.3% believed AI stakeholders would voluntarily follow EDI frameworks. As Chi et al. [15] and Cachat-Rosset and Klarsfeld [16] document, corporate AI ethics frameworks tend to redefine equity in narrowly technical terms parity in error rates, performance metrics which avoid confronting bias as a social and structural problem requiring institutional change.

Addressing this requires what Özbilgin et al. [14] call interdisciplinary co-design: the sustained integration of EDI scholars, social scientists, and community representatives alongside engineers throughout the development lifecycle, not merely as ethical reviewers of finished systems. The strong student support for structural interventions 81.4% endorsed mandatory pre-release equity audits and 69.1% supported EDI integration in university AI curricula signals that the normative foundation for such frameworks is already present among the next generation of practitioners, and that realizing this potential requires moving beyond compliance-oriented approaches toward genuine co-design where EDI principles are embedded from the outset [14], [16]. Policymakers and institutions should treat this as both a mandate and an opportunity. Adopting rigorous pre-deployment bias testing frameworks, analogous to the safety and efficacy phases of the pharmaceutical industry [2], and mandating publicly accessible bias audits across gender, race, geography, and occupation, represent concrete first steps toward governance structures commensurate with the scale and consequence of the bias documented here [7], [12].

Finally, 14.7% of open-ended respondents expressed "automation resignation" the view that AI bias is structurally inevitable and therefore not worth addressing. This represents a distinct challenge for EDI education. Students who internalize bias as a fixed feature of AI systems are less likely to become the advocates, developers, or policymakers needed to drive remediation [4], [24]. AI literacy curricula must therefore not only teach students to recognize bias but actively challenge the normative assumption that it is unavoidable, providing concrete evidence that representational outcomes are alterable through intentional design choices as the variation between 95% male-coded construction workers and 50% male-coded math teachers in this very dataset demonstrates [1], [14], [4], [22].

## VI. Limitations

While this study provides meaningful empirical insights, several limitations should be acknowledged. The survey sample was drawn exclusively from one institution (Mount Royal University) and was predominantly male (71%), which may limit the generalizability of the Phase 2 findings to broader student populations. Additionally, Phase 1 was conducted using a single AI model (GPT-3.5) at one point in time, meaning findings may not fully reflect the behavior of other large language models or more recent model versions. The gender coding of narratives, while systematic, relied on explicit linguistic markers and may not capture more subtle or implicit forms of gender representation embedded in narrative framing. These limitations do not undermine the core findings but rather point toward productive directions for future research, including replication across multiple institutions, AI platforms, and more demographically diverse samples.

## VII. Conclusion

This study sets out to examine whether gender bias emerges in AI-generated occupational narratives even when large language models are given gender-neutral prompts, and to explore whether the student population most actively using these systems is aware of, and trusting of, AI in consequential decision-making. Using a two-phase mixed-methods design, Phase 1 analyzed 360 ChatGPT-generated narratives across nine professions and four geographic regions, while Phase 2 surveyed 100 university students on their perceptions of AI bias, trust in AI systems, and attitudes toward EDI governance.

The findings from both phases converge on a common conclusion: gender bias in generative AI is a structurally embedded phenomenon that persists regardless of how neutrally a prompt is written, and it remains largely invisible to the everyday users most exposed to it. At the same time, students are not uncritical consumers of AI they recognize bias as a real problem, yet this awareness does not translate into trust when AI is applied to high-stakes decisions. Taken together, these findings confirm that addressing AI bias requires more than technical fixes; it demands interdisciplinary co-design that integrates EDI principles throughout the entire AI development lifecycle.

Future work should focus on building practical EDI-AI co-design frameworks that provide developers with concrete pre-deployment bias evaluation standards and equip students and educators with applied AI literacy tools that go beyond conceptual awareness. Replicating this study across multiple institutions, AI platforms, and more demographically diverse samples would strengthen the generalizability of these findings and bring the field closer to governance structures commensurate with the scale and consequence of the bias documented here.